\RequirePackage{lineno} 
\documentclass[prl,twocolumn,10pt,preprintnumbers,showpacs,showkeys,amsmath,amssymb,superscriptaddress]{revtex4}
\usepackage{graphicx}
\usepackage{dcolumn}
\usepackage{bm}
\usepackage{epsfig}
\usepackage[dvips]{color}
\begin{document}
\addtolength{\headheight}{2.0cm} 

\title{Studying Parton Energy Loss in Heavy-Ion Collisions via Direct-Photon and Charged-Particle Azimuthal Correlations}

\affiliation{Argonne National Laboratory, Argonne, Illinois 60439, USA}
\affiliation{University of Birmingham, Birmingham, United Kingdom}
\affiliation{Brookhaven National Laboratory, Upton, New York 11973, USA}
\affiliation{University of California, Berkeley, California 94720, USA}
\affiliation{University of California, Davis, California 95616, USA}
\affiliation{University of California, Los Angeles, California 90095, USA}
\affiliation{Universidade Estadual de Campinas, Sao Paulo, Brazil}
\affiliation{University of Illinois at Chicago, Chicago, Illinois 60607, USA}
\affiliation{Creighton University, Omaha, Nebraska 68178, USA}
\affiliation{Czech Technical University in Prague, FNSPE, Prague, 115 19, Czech Republic}
\affiliation{Nuclear Physics Institute AS CR, 250 68 \v{R}e\v{z}/Prague, Czech Republic}
\affiliation{University of Frankfurt, Frankfurt, Germany}
\affiliation{Institute of Physics, Bhubaneswar 751005, India}
\affiliation{Indian Institute of Technology, Mumbai, India}
\affiliation{Indiana University, Bloomington, Indiana 47408, USA}
\affiliation{University of Jammu, Jammu 180001, India}
\affiliation{Joint Institute for Nuclear Research, Dubna, 141 980, Russia}
\affiliation{Kent State University, Kent, Ohio 44242, USA}
\affiliation{University of Kentucky, Lexington, Kentucky, 40506-0055, USA}
\affiliation{Institute of Modern Physics, Lanzhou, China}
\affiliation{Lawrence Berkeley National Laboratory, Berkeley, California 94720, USA}
\affiliation{Massachusetts Institute of Technology, Cambridge, MA 02139-4307, USA}
\affiliation{Max-Planck-Institut f\"ur Physik, Munich, Germany}
\affiliation{Michigan State University, East Lansing, Michigan 48824, USA}
\affiliation{Moscow Engineering Physics Institute, Moscow Russia}
\affiliation{City College of New York, New York City, New York 10031, USA}
\affiliation{NIKHEF and Utrecht University, Amsterdam, The Netherlands}
\affiliation{Ohio State University, Columbus, Ohio 43210, USA}
\affiliation{Old Dominion University, Norfolk, VA, 23529, USA}
\affiliation{Panjab University, Chandigarh 160014, India}
\affiliation{Pennsylvania State University, University Park, Pennsylvania 16802, USA}
\affiliation{Institute of High Energy Physics, Protvino, Russia}
\affiliation{Purdue University, West Lafayette, Indiana 47907, USA}
\affiliation{Pusan National University, Pusan, Republic of Korea}
\affiliation{University of Rajasthan, Jaipur 302004, India}
\affiliation{Rice University, Houston, Texas 77251, USA}
\affiliation{Universidade de Sao Paulo, Sao Paulo, Brazil}
\affiliation{University of Science \& Technology of China, Hefei 230026, China}
\affiliation{Shandong University, Jinan, Shandong 250100, China}
\affiliation{Shanghai Institute of Applied Physics, Shanghai 201800, China}
\affiliation{SUBATECH, Nantes, France}
\affiliation{Texas A\&M University, College Station, Texas 77843, USA}
\affiliation{University of Texas, Austin, Texas 78712, USA}
\affiliation{Tsinghua University, Beijing 100084, China}
\affiliation{United States Naval Academy, Annapolis, MD 21402, USA}
\affiliation{Valparaiso University, Valparaiso, Indiana 46383, USA}
\affiliation{Variable Energy Cyclotron Centre, Kolkata 700064, India}
\affiliation{Warsaw University of Technology, Warsaw, Poland}
\affiliation{University of Washington, Seattle, Washington 98195, USA}
\affiliation{Wayne State University, Detroit, Michigan 48201, USA}
\affiliation{Institute of Particle Physics, CCNU (HZNU), Wuhan 430079, China}
\affiliation{Yale University, New Haven, Connecticut 06520, USA}
\affiliation{University of Zagreb, Zagreb, HR-10002, Croatia}

\author{B.~I.~Abelev}\affiliation{University of Illinois at Chicago, Chicago, Illinois 60607, USA}
\author{M.~M.~Aggarwal}\affiliation{Panjab University, Chandigarh 160014, India}
\author{Z.~Ahammed}\affiliation{Variable Energy Cyclotron Centre, Kolkata 700064, India}
\author{A.~V.~Alakhverdyants}\affiliation{Joint Institute for Nuclear Research, Dubna, 141 980, Russia}
\author{B.~D.~Anderson}\affiliation{Kent State University, Kent, Ohio 44242, USA}
\author{D.~Arkhipkin}\affiliation{Brookhaven National Laboratory, Upton, New York 11973, USA}
\author{G.~S.~Averichev}\affiliation{Joint Institute for Nuclear Research, Dubna, 141 980, Russia}
\author{J.~Balewski}\affiliation{Massachusetts Institute of Technology, Cambridge, MA 02139-4307, USA}
\author{L.~S.~Barnby}\affiliation{University of Birmingham, Birmingham, United Kingdom}
\author{S.~Baumgart}\affiliation{Yale University, New Haven, Connecticut 06520, USA}
% ...

\collaboration{STAR Collaboration}\noaffiliation

\date{\today}

\begin{abstract}
Charged-particle spectra associated with direct photon ($\gamma_{dir} $) 
and $\pi^0$ are measured in $p$+$p$ and Au+Au collisions at center-of-mass energy 
$\sqrt{s_{_{NN}}}=200$~GeV with the STAR detector at RHIC. A shower-shape analysis is 
used to partially discriminate between $\gamma_{dir}$ and $\pi^0$. Assuming no 
associated charged particles in the $\gamma_{dir}$ direction (near side) and 
small contribution from fragmentation photons ($\gamma_{frag}$), the associated 
charged-particle yields opposite to $\gamma_{dir}$ (away side) are extracted. 
At mid-rapidity ($|\eta|<0.9$) in central Au+Au collisions, charged-particle 
yields associated with $\gamma_{dir}$ and $\pi^0$ at high transverse momentum 
($8< p_{T}^{trig}<16$~GeV/$c$) are suppressed by a factor of 3-5 compared 
with $p$+$p$ collisions. The observed suppression of the associated charged 
particles, in the kinematic range $|\eta|<1$ and $3< p_{T}^{assoc} < 16$~GeV/$c$, 
is similar for $\gamma_{dir}$ and $\pi^0$, and independent of the $\gamma_{dir}$ 
energy within uncertainties. These measurements indicate that the parton energy 
loss, in the covered kinematic range, is insensitive to the parton path length.
\end{abstract}


\maketitle
\modulolinenumbers[5]
\linenumbers
A major goal of measurements at the Relativistic Heavy Ion Collider
(RHIC) is to quantify the properties of the QCD matter created in
heavy-ion collisions at high energy~\cite{STAR_whitepaper}. 
One key property is the
medium energy density, which can be probed
by its effect 
on a fast parton propagating through it~\cite{tomography}.
A parton scattered in the initial stages of a heavy-ion collision
propagates through the medium and results in a shower of hadrons (jet),
with high transverse momenta ($p_T$), in the detectors. 
The medium energy density is extracted through the comparison 
of measured observables with theoretical models. 
Many perturbative Quantum Chromodynamics (pQCD)-based models
of parton energy loss
have successfully described much of the high-$p_T$ data 
with medium parameters that span a wide range~\cite{Bass}.
To better constrain these
parameters, 
it is essential to examine the dependence of 
the energy loss ($\Delta E$) 
on the initial energy of the parton ($E$), 
path length of the parton through the 
medium ($L$), and the parton type, independently.
This necessitates additional experimental observables.

The $\gamma_{dir}$-jet coincidence measurements have long been proposed as a 
powerful tool to study parton energy loss  
in the medium created at RHIC~\cite{Wang_idea}. 
The leading-order production processes of direct photons $\gamma_{dir}$, 
quark-gluon Compton scattering $(q+g \rightarrow q+\gamma)$ and 
quark-antiquark annihilation $(q+\bar{q} \rightarrow g+\gamma)$, are free from the 
uncertainties accompanying fragmentation. The outgoing high-$p_{T}$ $\gamma$ balances 
the $p_T$ of the partner parton separated by $\pi$ in azimuth (``away-side''), modulo 
negligible corrections due to parton intrinsic $p_T$~\cite{Owens_Cormell}. 
The study of the spectra of the
away-side jet particles associated with the high-$p_{T}$ $\gamma_{dir}$ can 
constrain the dependence of $\Delta E$ on $E$.
The mean-free path of the $\gamma$ in the medium is large enough that its 
momentum is preserved, 
regardless of the position of the initial scattering vertex.
The $\gamma$ does not suffer from the geometric biases 
(non-uniform spatial sampling of hadron triggers due to energy loss in the medium) inherent in 
single hadron spectra and di-hadron azimuthal correlation measurements.
A comparison between the spectra 
of the away-side particles associated with $\gamma_{dir}$ vs. those associated 
with $\pi^{0}$ can constrain the dependence of
$\Delta E$ on $L$~\cite{Renk_dihadron}. 

In this Letter we examine $\Delta E$ by comparing jet yields measured in
central Au+Au collisions and $p$+$p$ collisions at $\sqrt{s_{_{NN}}} = 200$~GeV via
correlations in azimuthal angle between particles. 
We investigate the $\Delta E$ dependence on $E$, via $\gamma_{dir}$-charged-particle 
($\gamma_{dir}$-$h^{\pm}$) correlations, and on
$L$, via a comparison of   
$\pi^{0}$-charged particle ($\pi^{0}$-$h^{\pm}$) to ($\gamma_{dir}$-$h^{\pm}$) 
correlations.
Taking advantage of the unique configuration of the STAR detector,
we present a novel analysis technique to extract the spectra of charged 
particles associated with $\gamma_{dir}$. 
This technique provides a much needed higher 
statistics than that in ~\cite{PHENIX_gjet} for these types of rare probes, and
allows for more statistically significant measurement. 

The STAR detector is well suited for measuring azimuthal angular correlations 
due to the large coverage in pseudorapidity ($\eta$)
and full coverage in azimuth ($\phi$). Using the Barrel Electromagnetic
Calorimeter (BEMC)~\cite{STAR_BEMC} to select events (\textit{i.e.} ``trigger") with high-$p_{T}$ $\gamma$, the
STAR experiment collected an integrated 
luminosity of 535~$\mu$b$^{-1}$ of Au+Au 
collisions in 2007 and 11~pb$^{-1}$ of $p$+$p$ collisions in 2006. 
The BEMC consists of 4800 channels (towers) and measures the 
$\gamma$ energy. 
The Time Projection Chamber (TPC)~\cite{STAR_TPC} detects 
charged-particle tracks.
A crucial part of the analysis is to discriminate between showers from $\gamma_{dir}$ and 
two close $\gamma$'s from high-$p_{T}$ $\pi^{0}$ symmetric decays. 
At $p_T^{\pi^0} \sim 8$~GeV/$c$, the angular separation between the two $\gamma$'s 
resulting from a $\pi^{0}$ decay is typically smaller than a tower 
size, but a $\pi^{0}$ 
shower is generally broader than a single $\gamma$ shower. The Barrel Shower Maximum 
Detector (BSMD)~\cite{STAR_BEMC} consists of 18000 channels (strips) 
in each plane ($\eta$ and $\phi$) and resides at $\sim 5.6$  
radiation lengths inside the calorimeter towers. The BSMD is capable of 
$(2\gamma$)/$(1\gamma)$ separation up to $p_T^{\pi^0} \sim 26$~GeV/$c$ due 
to its high granularity. 

In this analysis, events with vertex within $\pm 55$ cm 
of the center of TPC are selected. 
The BEMC is calibrated using the 2006 $p$+$p$ data, 
using a procedure described elsewhere~\cite{STAR_pp_pi0}. 
The tracking efficiency of charged 
particles as a function of event multiplicity 
is determined by embedding $\pi^{\pm}$ in real data.
The effects of energy and momentum resolution are estimated 
to be small compared to other systematic uncertainties in this 
analysis, and no correction is applied.
The charged-track quality criteria are similar to those 
used in previous STAR analyses~\cite{Magestro}. 
Events with at least one electromagnetic cluster 
(defined as 1 or 2 towers) with $E_T > 8$~GeV are selected. 
A trigger tower is rejected if it has a track 
with $p > 3.0 $~GeV/$c$ pointing to it.
The $\pi^{0}$/$\gamma$ discrimination depends on an analysis 
of the shower shape as measured by the BSMD and BEMC. 
The shower shape is studied by single-particle Monte-Carlo 
simulation and embedding $\gamma$ and $\pi^{0}$ in real data. 
The shower shape is quantified with the cluster energy, 
measured by the BEMC, normalized by the position-dependent energy moment, 
measured by the BSMD strips.
The shower profile cuts were tuned to obtain a $\gamma_{dir}$-free 
($\pi^{0}_{rich}$) sample and a sample rich in $\gamma_{dir}$ ($\gamma_{rich}$). 
From embedding single $\gamma$'s and $\pi^0$'s into Au+Au data,
the rejection power of the shower profile cuts is estimated to be $>99$\% for 
rejecting $\gamma_{dir}$ and $\sim 60$\% for rejecting $\pi^0$.
A detailed study of the shower profile, primary vertex, and charge-rejection cuts is 
performed to determine the systematic uncertainties, which also include the energy scale uncertainty.
\begin{figure}
   \resizebox{90mm}{!}{\includegraphics{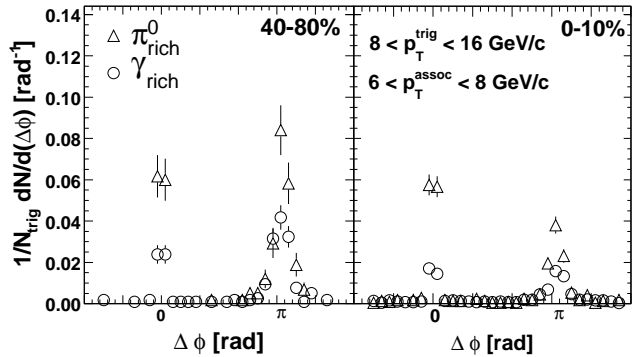}} 
        \caption{Correlations (without background subtraction) of $\gamma_{rich}$ sample of triggers 
    and $\pi^{0}_{rich}$ sample with associated charged hadrons in peripheral (40-80\%) 
    and central (0-10\%) Au+Au collisions.}
    \label{fig:corrfunc}
\end{figure} 

The correlations in $\Delta\phi$ between particles, measured as the number
of associated particles per trigger per $\Delta\phi$ (``correlation functions''), 
are used in both $p$+$p$ and 
Au+Au collisions to determine the (jet) associated particle yields.
Figure~\ref{fig:corrfunc}   
shows the correlation functions for
$\gamma_{rich}$ and $\pi^{0}_{rich}$ triggers for the peripheral (largest impact parameters) and 
most central (smallest impact parameters) 
bins in Au+Au collisions.
As expected the 
$\gamma_{rich}$-triggered sample has lower near-side yields than 
those of the $\pi^{0}_{rich}$, but not zero. In addition to the isotropically distributed underlying-event background, 
the non-zero near-side yield for the $\gamma_{rich}$ sample is expected due 
to remaining background contributions of widely separated $\gamma$'s that are correlated with charged particles. 
The shower-shape analysis 
is only effective for rejecting two close $\gamma$ showers, 
leaving background $\gamma$'s from
asymmetric decays of $\pi^{0}$ and $\eta$, and
fragmentation $\gamma$'s.

The uncorrelated background level is subtracted, assuming an isotropic 
distribution determined by fitting the correlation function
with two Gaussians and a constant. Over the measured range of $p_{T}^{assoc}$ 
the expected modulation in the background shape, 
due to the correlation with respect to the reaction plane in heavy-ion collisions, 
is found to have a negligible effect on the subtraction.
As shown in Fig. ~\ref{fig:corrfunc}, the level of uncorrelated background 
is dramatically suppressed relative to the signal.
The near- and away-side yields, $Y^{n}$ and $Y^{a}$, of associated particles per trigger are extracted by 
integrating the $\mathrm 1/N_{trig} dN/d(\Delta\phi)$ distributions, over $\mid\Delta\eta\mid <1.9$, in $\mid\Delta\phi\mid$ $\leq$~0.63 
and $\mid\Delta\phi -\pi\mid$  $\leq$~0.63, respectively. The yield is corrected for the tracking efficiency 
of charged particles as a function of event multiplicity but, as in~\cite{Magestro}, not for acceptance due to the $\eta$ cuts.

\begin{figure}
   \resizebox{85mm}{!}{\includegraphics{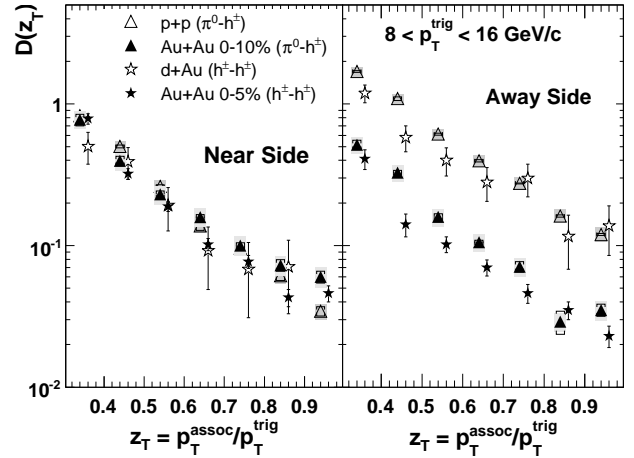}} 
         \caption{The $z_{T}$ dependence of $\pi^{0}-h^{\pm}$ and $h^{\pm}-h^{\pm}$~\cite{Magestro} near-side (left panel) and away-side (right panel) 
    associated particle yields. The bin centers are shifted for clarity.  The shaded boxes show the systematic errors correlated in $z_{T}$, and the brackets show the point-to-point systematic errors.}
     \label{fig:pi0} 
\end{figure}

Figure~\ref{fig:pi0} shows the
hadron yields associated with $\pi^{0}_{rich}$ 
normalized by the measured number 
of triggers ($D(z_T)$~\cite{Wang_idea}), as a function of
$z_{T}= p_{T}^{assoc}/p_{T}^{trig}$, 
compared to the yields per charged-hadron trigger~\cite{Magestro}. 
The yield in the first $z_T$ bin is corrected for the $\Delta z_T$ width sampled 
on a trigger-by-trigger basis.
The systematic errors on the $\pi^0_{rich}$-triggered yields have 
a correlated component of $\sim 7-13$\%, 
and point-to-point uncertainties that are less than 5\% for much of the data.
Since the charged-hadron triggers are dominated by charged pions,
the associated yields are expected to be similar to those of $\pi^0$ triggers, although there could be some
differences due to the presence of proton triggers in the charged-trigger sample.
A general agreement of $\sim 20-30$\% between the results from this analysis ($\pi^{0}-h^{\pm}$) and the previous STAR analysis 
($h^{\pm}-h^{\pm}$) is clearly seen in both panels of Fig. 2,
indicating the $\pi^{0}_{rich}$-sample is free of $\gamma_{dir}$.

Assuming zero near-side yield for $\gamma_{dir}$ triggers and a  
sample of $\pi^{0}$ free of $\gamma_{dir}$, 
the away-side yield of hadrons correlated with the $\gamma_{dir}$ is 
extracted as
\begin{align}
Y_{\gamma_{dir}+h}&=\frac{Y^{a}_{\gamma_{rich}+h}-{\cal{R}} Y^{n}_{\gamma_{rich}+h}}{1-r}, 
{\rm where~} {\cal{R}}=\frac{Y^{a}_{\pi^{0}_{rich}+h}}{Y^{n}_{\pi^{0}_{rich}+h}},\nonumber \\
r&=\frac{Y^{n}_{\gamma_{rich}+h}}{Y^{n}_{\pi^{0}_{rich}+h}}, 
{\rm ~~~and~} 1-r=\frac{N^{\gamma_{dir}}}{N^{\gamma_{rich}}}. 
\label{eq:one}
\end{align}
Here, $Y^{a(n)}_{\gamma_{rich}+h}$ and $Y^{a(n)}_{\pi^{0}_{rich}+h}$ are 
the away (near)-side yields of associated particles 
per $\gamma_{rich}$ and $\pi^{0}_{rich}$ triggers, respectively.  The 
ratio $r$ is equivalent to the fraction of ``background'' triggers 
in the $\gamma_{rich}$ trigger sample, and 
$N^{\gamma_{dir}}$ and $N^{\gamma_{rich}}$ are the numbers of $\gamma_{dir}$ and $\gamma_{rich}$ triggers, respectively. The value or $r$
is found to be $\sim 55\%$ in $p$+$p$ and decreases to $\sim 30\%$ in central Au+Au with little dependence on $p_{T}^{trig}$. 
All background to $\gamma_{dir}$ is subtracted with the assumption that 
the background triggers have the same correlation function
as the $\pi^{0}_{rich}$ sample.
\textsc{pythia} simulations~\cite{pythia} indicate that correlations of 
$\gamma$ triggers from asymmetric hadron decays 
are similar, to within $\sim 10$\%, 
to those of symmetrically decaying $\pi^0$ triggers 
as well as the measured correlations of $\pi^0_{rich}$ triggers, 
at the same $p_T^{trig}$. 
On the other hand, \textsc{pythia} shows that the $\gamma_{frag}$ has a different 
correlation with the charged particle compared to that of $\pi^0$ with insensitivity to the charged rejection cut. 
However, 
the $\gamma_{frag}$ contribution is expected to fall off more rapidly in 
$x_{T}$ ($x_{T} = 2p_{T}/\sqrt{s}$) than the other lowest
order $\gamma_{dir}$'s~\cite{photons_theory}.
One theoretical
calculation~\cite{Frag_photons} shows the ratio of $\gamma_{frag}$ to $\gamma_{dir}$ 
to be $\sim 30-40$\% at $p_{T}^{\gamma} > 8$~GeV/c in $p$+$p$ at mid-rapidity at RHIC energy. 

For the $\gamma_{dir}$-triggered yields, the systematic errors are 
evaluated similar to $\pi^{0}$ and
summarized 
as a function of centrality and $z_T$ in Table~\ref{table:syserrors}.
\begin{table}[h]
\caption{Systematic errors on $\gamma_{dir}$-triggered yields.}
\begin{tabular}{c c}
\hline
~~Au+Au ~0-10\% collisions & $z_T$-correlated error: 17-19\% ~~\\
\end{tabular}
\begin{tabular}{c | c c c c c c}
\hline
$z_T$ bin & 0.35 & 0.45 & 0.55 & 0.65 & 0.75 & 0.85 \\
point-to-point error (\%) &  37  & 21  & 21 & 55 & 20 &  49 \\
\hline
\end{tabular}
\begin{tabular}{c c}
pp collisions ~~ & $z_T$-correlated error: 11-13\% \\
\end{tabular}
\begin{tabular}{c | c c c c c c}
\hline
$z_T$ bin & 0.35 & 0.45 & 0.55 & 0.65 & 0.75 & 0.85 \\
point-to-point error (\%) & 10  & 13 &  16 & 93 & 17 &  24  \\
\hline
\end{tabular}
\label{table:syserrors}
\end{table}
An additional 
source of uncertainty on these yields arises from 
the assumption that the background contribution of $\gamma_{frag}$
in the $\gamma_{rich}$ triggers
has the same correlation as the $\pi^0_{rich}$ triggers. 
This is assessed 
by comparing (with a $\chi^2$ analysis) the shape of 
the near-side correlation of $\gamma_{rich}$ to $\pi^0_{rich}$ triggers.
Thus, excepting those $\gamma_{frag}$ that have no near-side yield,
the contribution of $\gamma_{frag}$ in the
$\gamma_{rich}$ sample, not satisfying the assumption, is taken
into account in the systematic errors.
\begin{figure}
   \resizebox{70mm}{!}{\includegraphics{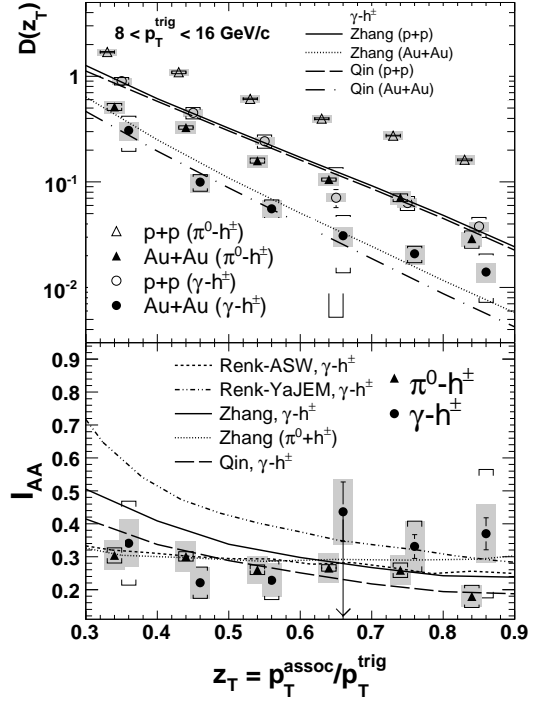}} 
      \caption{Upper panel: $z_{T}$ dependence of away-side associated-particle yields for $\pi^{0}$ triggers (triangles) and $\gamma_{dir}$ triggers (circles) for $p$+$p$ (open symbols) and
    0-10\% Au+Au (closed symbols) collisions.   The trigger particle has $8 < E_T^{trig} < 16$~GeV/$c$.  
    Lower panel: $z_{T}$ dependence of $I_{AA}$ for $\gamma_{dir}$ triggers (circles) and $\pi^{0}$ triggers (triangles).
    Boxes
    show the $z_T$-correlated systematic errors, and brackets show the point-to-point systematic errors. The bin centers are shifted for clarity.  Data is compared to theoretical calculations (see text).}
   \label{fig:gamma_zt}
\end{figure}

Figure~\ref{fig:gamma_zt} (upper panel) shows the $z_{T}$ dependence 
of the trigger-normalized fragmentation function for 
($\pi^{0}-h^{\pm}$) and 
($\gamma-h^{\pm}$) in $p$+$p$, and 0-10\% central Au+Au collisions. 
At a given $z_{T}$, the away-side yield per $\pi^{0}$ trigger is significantly larger 
than the yield per $\gamma_{dir}$ trigger. 
This difference is expected,
since the $\gamma_{dir}$ carries the total 
scattered constituent momentum while the $\pi^{0}$ carries only a fraction of it.  In addition, there are
different proportions of quarks and gluons recoiling from   
$\gamma_{dir}$ and $\pi^{0}$ triggers. 
In Au+Au collisions, partonic energy loss can lead to additional differences 
on the away-side since the path length, the energy of the parton, and
the partonic species composition of the recoiling parton are different between
$\gamma_{dir}$ and $\pi^0$ triggers at the same $p_T^{trig}$.
A comparison to two different theoretical calculations of the 
associated yields for $\gamma_{dir}$ triggers is shown 
in Fig.~\ref{fig:gamma_zt} (upper panel). 
The calculation by Zhang~\cite{Wang_gamma} 
does not include $\gamma_{frag}$ and describes the data well.
The calculation by Qin~\cite{Gale_gamma} includes a significant contribution of $\gamma_{frag}$, 
but it is quite similar in
yield for p+p and also describes the data within current uncertainties.

\begin{figure}
   \resizebox{60mm}{!}{\includegraphics{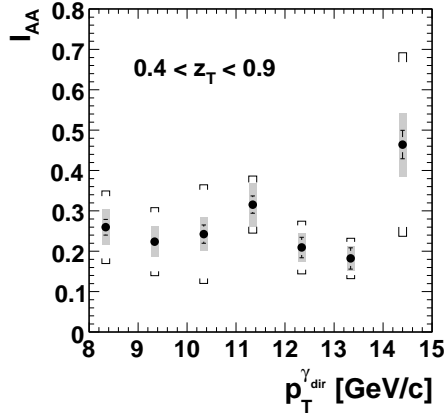}} 
        \vspace{-0.2cm}
    \caption{$I_{AA}$ as a function of $p_{T}$ for $\gamma_{dir}$ triggers, measured in 0-10\% Au+Au collisions.  The associated charged particles have $z_T = 0.4-0.9$.
The shaded boxes show the systematic errors correlated in $p_{T}^{trig}$, and the brackets show the point-to-point systematic errors.}
       \label{fig:pttrig} 
   \vspace{-0.2cm}
\end{figure}

In order to quantify the away-side suppression, 
we calculate the quantity $I_{AA}$, which is defined as the ratio of the integrated yield of the away-side associated 
particles per trigger particle in Au+Au to that in $p$+$p$ collisions. 
Figures~\ref{fig:gamma_zt} (lower panel) and~\ref{fig:pttrig}
show $I_{AA}$ as a function of $z_T$ and $p_{T}^{trig}$, respectively.
Despite the differences between $\pi^{0}-h^{\pm}$ and $\gamma-h^{\pm}$, seen in Fig.~\ref{fig:gamma_zt} (upper
panel), 
the value of $I_{AA}^{\gamma-h^{\pm}}$ is $z_{T}$ independent and similar to that of 
$I_{AA}^{\pi^{0}-h^{\pm}}$. The $I_{AA}^{\gamma-h^{\pm}}$ agrees well with 
theoretical calculations in which the energy loss is tuned to the single- and di-hadron 
measurements~\cite{RAA_pi0,Magestro}.  The calculation by Zhang for both $\gamma_{dir}$ and $\pi^0$
triggers~\cite{Wang_gamma} shows only 
a small difference in the suppression factor, growing at low $z_T$.  Two calculations for $\gamma_{dir}$ triggers, 
labeled as Qin~\cite{Gale_gamma} and Renk-ASW~\cite{Renk_gamma},
show even less of a rise at low $z_T$.
In the calculation~\cite{Renk_gamma} using the ASW implementation of energy loss~\cite{ASW}, the effect of 
fluctuations in energy loss dominates over the effect of geometry, explaining the similarity
in $\gamma$- and $\pi^0$-associated yields.  
The calculation that is not consistent with the data at low $z_T$, 
the Renk-YaJEM model~\cite{Renk_gamma}, differs in that the lost energy is tracked and redistributed through the medium.
The disagreement with this model may indicate that the lost energy is distributed to extremely low $p_T$ and 
large angles~\cite{Renk_gamma} (as also evidenced by hadron-hadron correlation measurements~\cite{AS_cone}), and
perhaps even that the correlations to the trigger particle are lost.
To further test this, one must explore the region of low $z_{T}$. 

Figure~\ref{fig:pttrig} addresses the $E$ dependence of $\Delta E$.
The suppression of the away-side multiplicity per $\gamma_{dir}$ 
trigger in Au+Au relative to $p$+$p$ collisions
shows no strong $p_{T}^{trig}$ dependence, 
which indicates no strong $E$ dependence in the measured
$p_T$ range.

In summary, $\gamma_{dir}-h$ correlation measurements
are reported by the STAR collaboration, providing important new constraints on
theoretical models.
The agreement between the measured $I_{AA}^{\gamma-h^{\pm}}$ and 
$I_{AA}^{\pi^{0}-h^{\pm}}$ 
in the covered kinematic range
could theoretically be due to an interplay of compensating
factors of the dependence of energy loss on the initial parton energy, 
the energy loss of gluons vs. quarks, 
and the energy loss path-length dependence. 
The measurement of the dependence of $I_{AA}^{\gamma-h^{\pm}}$
on $p_T^{trig}$, however, shows no significant dependence on the 
initial parton energy. Therefore, 
the dependence of observable parton energy loss on parton species and path length 
traversed by the parton in the medium
must be similarly small.

We thank the RHIC Operations Group and RCF at BNL, the NERSC Center 
at LBNL and the Open Science Grid consortium for providing resources and support. 
This work was supported in part by the Offices of NP and HEP within the U.S. DOE Office of Science, 
the U.S. NSF, the Sloan Foundation, the DFG cluster of excellence `Origin and Structure of the Universe', 
CNRS/IN2P3, STFC and EPSRC of the United Kingdom, FAPESP CNPq of Brazil, Ministry of Ed. and Sci. of the Russian Federation, 
NNSFC, CAS, MoST, and MoE of China, GA and MSMT of the Czech Republic, FOM and NWO of the Netherlands, DAE, DST, 
and CSIR of India, Polish Ministry of Sci. and Higher Ed., Korea Research Foundation, Ministry of Sci., Ed. 
and Sports of the Rep. Of Croatia, Russian Ministry of Sci. and Tech, and RosAtom of Russia.

\end{document}